\documentstyle[psfig,sprocl]{article}
\def\Journal#1#2#3#4{{#1} {\bf #2}, #3 (#4)}
\def\msun{$M_{\odot}$}

\def\etal{{\it et al. }}

\def\be{\begin{equation}}
\def\ee{\end{equation}}
\def\bea{\begin{eqnarray}}
\def\eea{\end{eqnarray}}

\begin{document}
\title{Multifrequency Observations of the Galactic Microquasars GRS1915+105 and GROJ1655-40}
\author{Ronald A. Remillard, Edward H. Morgan (MIT)}
\author{Jeffrey E. McClintock (CFA), Charles D. Bailyn (Yale)}
\author{Jerome A. Orosz (Penn State) \& Jochen Greiner (AIP, Potsdam)}
\maketitle\abstracts { The two galactic `microquasars' with
superluminal radio jets have been quite active during 1996, generating
a variety of studies involving both NASA and ground-based
observatories.  GRS~1915+105 has displayed dramatic accretion
instability in observations with RXTE, revealing X-ray light curves
and emission states unlike anything previously seen.  Variable QPOs in
the range of 0.07--10 Hz have been monitored with the capability to
track the individual oscillations.  The QPO amplitude is as high as
40\% of the mean flux, while both amplitude and phase lag increase
with photon energy. The results imply a direct link between the QPO
mechanism and the origin of the energetic electrons believed to
radiate the X-ray power-law component.  GRS1915+105 also displays a
transient yet stationary QPO at 67 Hz. The other source, GRO J1655-40,
is an optically established black hole binary.  Recent optical reports
include an excellent model for the binary inclination and masses,
while an optical precursor to the April 1996 X-ray outburst has been
measured.  We report new results from recent RXTE observations.
GROJ1655-40 has displayed both the canonical ``soft/high'' and ``very
high'' X-ray states, with QPOs at 8--22 Hz during the latter state. In
addition, there is a high-frequency QPO at 300 Hz.  The rapid
oscillations in these sources are suspected of providing a measure of
the mass and rotation of the accreting black holes, although several
competing models may be applied when evaluating the results. }

\section{Introduction}
	The two sources of superluminal radio
jets~\cite{mirrod}~\cite{tin}~\cite{hj95} in the Galaxy, GRS1915+105
and GRO J1655-40 have been quite active during 1996.  These X-ray
sources were originally detected during May 1992 and July 1994,
respectively, and they have persisted well beyond the typical time
scale for X-ray transients~\cite{clg}.  Optical study of the companion
star in GROJ1655-40 has yielded a binary mass function (3.2 \msun)
that indicates an accreting black hole~\cite{cb}~\cite{ob}. In the
case of GRS1915+105, interstellar extinction limits optical/IR studies
to weak detections at wavelengths $> 1$ micron~\cite{mir94}.  The
compact object in this system is supsected of being a black hole due
to the spectral and temporal similarities with GROJ1655-40 and other
black hole binaries. Both of these microquasars have now been detected
with OSSE~\cite{g97} out to photon energies of 600 keV.

 	Investigations of microquasars are motivated by several broad
and interrelated purposes: to search for clues regarding the origin of
relativistic jets, to probe the properties of the compact objects, and
to understand the various spectral components and their evolution as
the sources journey through different accretion states.  Several
research programs are described herein, with emphasis on new results
from the Rossi X-ray Timing Explorer (RXTE).

\section{RXTE Observations of GRS1915+105}
	The RXTE All Sky Monitor~\cite{lev96} began operation during
1996 Jan 5-13, and continuous observing with a 40\% duty cycle has
been achieved since 1996 Feb 20.  GRS1915+105 was found to be bright
and incredibly active~\cite{mor}, as ASM time series data revealed
high amplitude modulations at 10-50 s. These results initiated a
series of weekly pointings for the PCA and HEXTE instruments.  The
yield is approaching ten billion photons in an immensely complex and
exciting archive that is fully available as `public' data.

	The ASM light curve of GRS1915+105 (1996 Feb 20 -- 1997 Jan
23) is shown in Fig. 1.  These results are derived using version 2
(1/97) of the model for the instrumental response to X-ray shadows
through the coded masks. The top panel shows the normalized intensity
for the full range (2--12 keV) of the ASM cameras, in which the Crab
nebula produces 75.5 c/s. The vertical lines in the upper region show
the times of the PCA / HEXTE observations in the public archive.
Below this light curve, one of the ASM hardness ratios is displayed;
$HR2$ is the ratio of normalized flux at 5--12 keV relative to the
flux in the 3--5 keV band.  The spectrum of GRS1915+105 is harder than
the Crab ($HR2$ = 1.07). Since there is an anticorrelation between the
count rate and $HR2$ in GRS1915+105, we caution against the
presumption that the ASM flux is a direct measure of X-ray luminosity.
During 1997, significant progress is expected from efforts to combine
the ASM results with those of BATSE and radio monitors, including the
newly organized Greenbank Interferometer project.  This effort will
build on earlier work~\cite{h95} to investigate the multifrequency
evolution of X-ray outbursts and radio flares.

	The PCA observations of GRS1915+105 immediately showed
dramatic intensity variations~\cite{gmr} with a complex hierarchy of
quasi-periodic dips on time scales from 10 s to hours.  Complex and
yet repeatable `stalls' in the light curve were preceeded by rapid
dips in which the count rate dropped by as much as 90\% in a few
seconds. These variations were interpreted as an inherent accretion
instability, rather than absorption effects, since there was spectral
softening during these dips.  There were also occasions of flux
overshooting after X-ray stalls.  These repetitive, sharp variations
and their hierarchy of time scales are entirely unrelated to the
phenomenology of absorption dips~\cite{gmr}.  The dips represent large
changes in an absolute sense; the pre-dip or post-dip luminosity in
GRS1915+105 is as high as $2\times 10^{39} ~{\rm ergs}~{\rm
cm}^{-2}~{\rm s}^{-1}$ at 2-60 keV, assuming the distance of 12.5 kpc
inferred from 21 cm HI absorption profiles~\cite{mirrod}.

	The phenomenology of wild source behavior in GRS1915+105 has
expanded since the first series of observations. Three examples are
shown in Fig. 2. The Oct 7 display of quasiperiodic stalls preceeded
by rapid dips (middle panel) is highly organized and repetitive, while
the Jun 16 light curve (top panel) shows complex, interrupted stalls
that are not preceeded by rapid dips.  In the bottom panel, an entirely new
type of oscillatory instability is displayed; hundreds of these
ringing features were recorded during Oct 13 and 15 with a recurrence
time near 70 s.  During Oct 15 the recurrence time increases
(see Fig. 2), leading to a long X-ray stall and subsequent flux
overshoot. The nature of these astonishing X-ray instabilities is
currently a mystery.  Note, however, that most of the PCA observations
show `normal' light curves with variations limited to rapid flickering
at 10-20 \% of the mean rate.

	A penetrating analysis of GRS1915+105 was made by
investigating the X-ray power spectra and comparing them with the
characteristics of the ASM light curve~\cite{mrg}.  The shape of the
broad-band power continuum and the properites of rapid QPOs (0.01 to
10 Hz) are correlated with the brightness, spectral hardness, and the
long-term variations seen with the ASM. Four emission states were
found, labelled in Fig. 1 as chaotic (CH), bright (B), flaring (FL),
and low-hard (LH). We see QPOs and nonthermal spectral components
during all four states, implying that they are new variants of the
`very high state' rarely seen in other X-ray
binaries~\cite{vdk94}~\cite{vdk96}.  The combination of the intense
QPOs and the high throughput of the PCA enabled phase tracking of
individual oscillations. Four QPO cases were chosen from three
different states~\cite{mrg}, with frequencies ranging from 0.07 to 2.0
Hz. The results are remarkably similar: the QPO arrival phase
(relative to the mean frequency) exhibits a random walk with no
correlation between the amplitude and the time between subsequent
events.  Furthermore the mean `QPO-folded' profiles are roughly
sinusoidal with increased amplitude at higher energy, and with a
distinct phase lag of $\approx$ 0.03 between 3 and 15 keV.  At photon
energies above 10 keV, the high amplitudes and sharp profiles of the
QPOs are inconsistent with any scenario in which the phase delay is
caused by scattering effects. Alternatively, it appears that the
origin of the hard X-ray spectrum itself (i.e. the creation of
energetic electrons in the inverse Compton model) is functioning in a
quasiperiodic manner. These results fundamentally link X-ray QPOs with
the most luminous component of the X-ray spectrum in GRS1915+105.

	In addition to the frequent X-ray QPOs below 10 Hz, a
transient yet `stationary' QPO at 67 Hz has been
discovered~\cite{mrg}.  This feature is seen on 6 of the first 31 PCA
observations of GRS1915+105. Typically, the amplitude is 1\% of the
flux and the QPO width is 3.5 Hz. This QPO exhibits a strong energy
dependence, rising (e.g. on 1996 May 6) from 1.5 \% at 3 keV to 6\% at
15 keV. One may attempt to associate this frequency with the mass and
spin rate of an accreting black hole, but the competing models include
such concepts as instabilities at the minimum stable orbit of $3 Rs$,
implying a mass of 33 \msun ~for a nonrotating black hole~\cite{mrg},
to relativistic modes of oscillation in the inner accretion disk,
implying 10 \msun ~for a nonrotating black hole~\cite{now}.

\section{Recent Observations of GRO J1655-40}

	During much of 1995 and early 1996, GRO J1655-40 was in a low
or quiescent accretion state, permitting a clear optical view of the
companion star (near F4 IV). Orosz and Bailyn~\cite{ob} improved the
determinations of the binary period (2.62157 days) and the mass
function.  They further measured the `ellipsoidal variations' arising
from the rotation of the gravitationally distorted companion star.
Their analysis, using B,V,R, and I bandpasses, provide an
exceptionally good fit for the binary inclination angle (69.5 deg) and
the mass ratio.  From these results, they deduce masses of $7.0 \pm
0.2$ and $2.34 \pm 0.12$ \msun ~for the black hole and companion star,
respectively.
	
	The ASM recorded a renewed outburst from GRO J1655-40 that
began on 1996 April 25. The ASM light curve (Feb 1996 to Jan 1997) is
shown in the lower half of Fig. 1. With great fortune, our optical
campaign had lasted until April 24, and Orosz \etal has
shown~\cite{ob2} that optical brightening preceeded the X-ray ascent
by 6 days, beginning first in the I band and then accelerating quickly
in blue light.  These results provide concrete evidence favoring the
accretion disk instability as the cause of the X-ray nova.  Theorists
may now attempt to model the brightness gradients and delay times in
the effort to develop a deeper understanding of this outburst.

	The ASM $HR2$ measures (Fig. 1) show an initially soft
spectrum that becomes brighter and harder for several months during
mid outburst.  The PCA observations from our GO program confirm this
evolution, as the power-law component (photon index $\approx 2.6$)
dominates the spectrum during the brightest cases.  The great majority
of PCA measurments of GRO J1655-40 follow single tracks on the
intensity:color and color:color diagrams, with a positive correlation
between hardness and brightness.

	PCA power spectra show transient QPOs in the range of 8--22 Hz
that are clearly associated with the strength of the power-law
component.  Using a PCA-based hardness ratio, $PCA\_HR2$ = flux above
9.6 keV / flux at 5.2--7.0 keV, we detect QPO in the range of 8--22 Hz
whenever $PCA\_HR2 > 0.22$.  Furthermore, in the 7 `hardest'
observations ($PCA\_HR2 > 0.3$), there is evidence of a high-frequency
QPO near 300 Hz. In Fig. 3 we show the sum of PCA power spectra in
these 3 intervals of $PCA\_HR2$, illustrating the QPO centered at 298
Hz. The Poisson noise has been subtracted, with inclusion of deadtime
effects~\cite{mrg}. The integrated feature has a significance of $14
\sigma$, a width of 120 Hz, and an amplitude near 0.8\%.  Applying the
`last stable orbit' model to this feature yields a mass of 7.4 \msun
~for a non-rotating black hole.  While this is astonishingly similar
to the optically determined mass, we caution that other models can
give similar results in the case of significant black hole rotation.
We further note that none of the models discussed~\cite{mrg} for the
high-frequency QPOs in GROJ1655-40 and GRS1915+105 adequately address
the spectral signature of this oscillation, which is more directly
associated with the power law component rather than the disk (thermal)
component. Nevertheless, the fact of these QPOs, which almost
certainly originate very near the accreting compact objects, will
remain a vigorous research topic throughout the RXTE Mission.

\begin{figure}
\psfig{figure=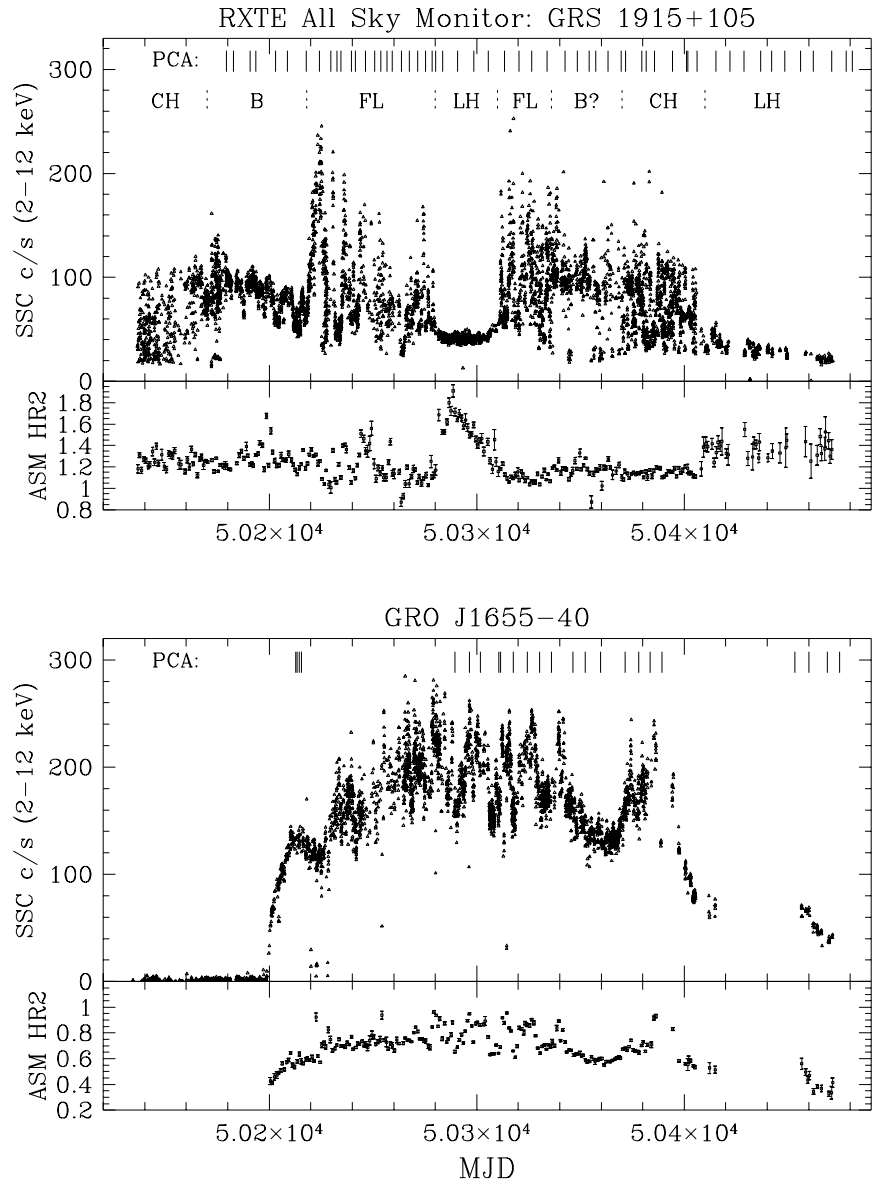,height=5.5in}
\vskip 1.5cm
\caption{RXTE ASM light curves and hardness ratio of GRS1915+105 and GRO J1655-40.
\label{fig:asm19}}
\end{figure}

\begin{figure}
\psfig{figure=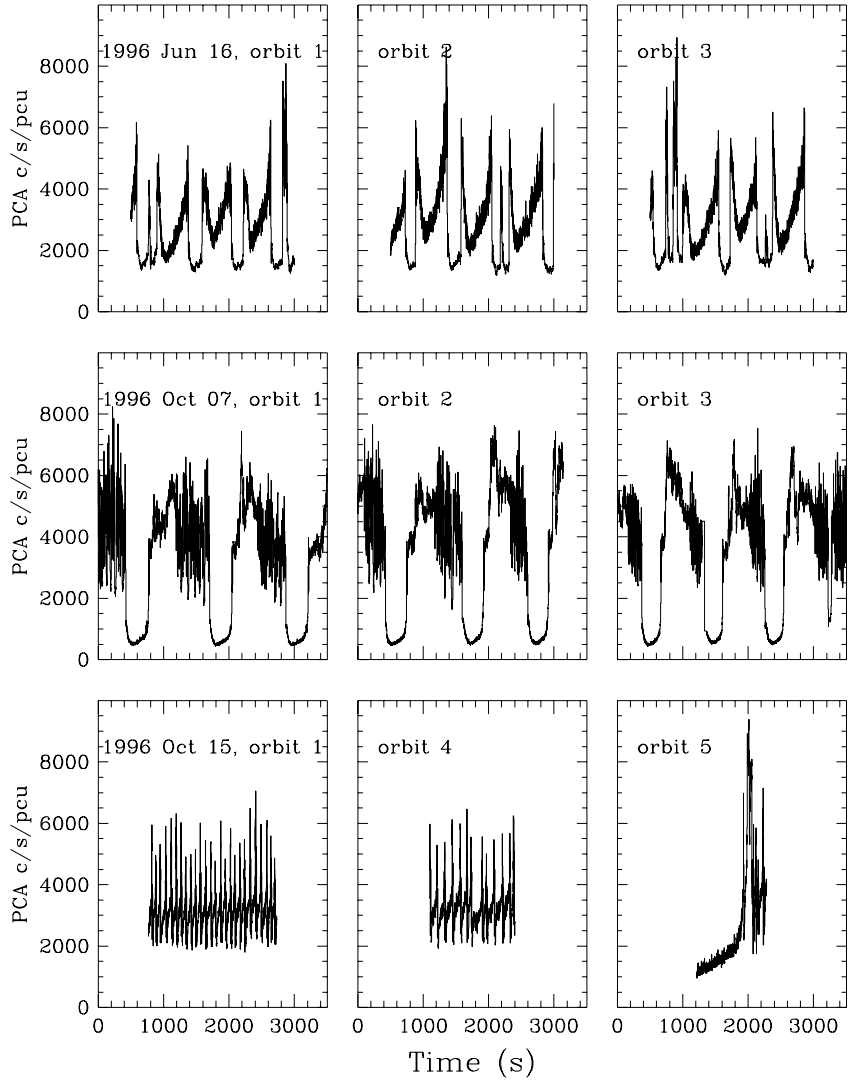,height=5.25in} 
\vskip 1.0cm
\caption{Samples of PCA light curves showing dramatic variability in
GRS1915+105 (2--30 keV). The adjacent panels along each row display
source measurements from different satellite orbits during the same
observation.
\label{fig:pca19}}
\end{figure}

\begin{figure}
\psfig{figure=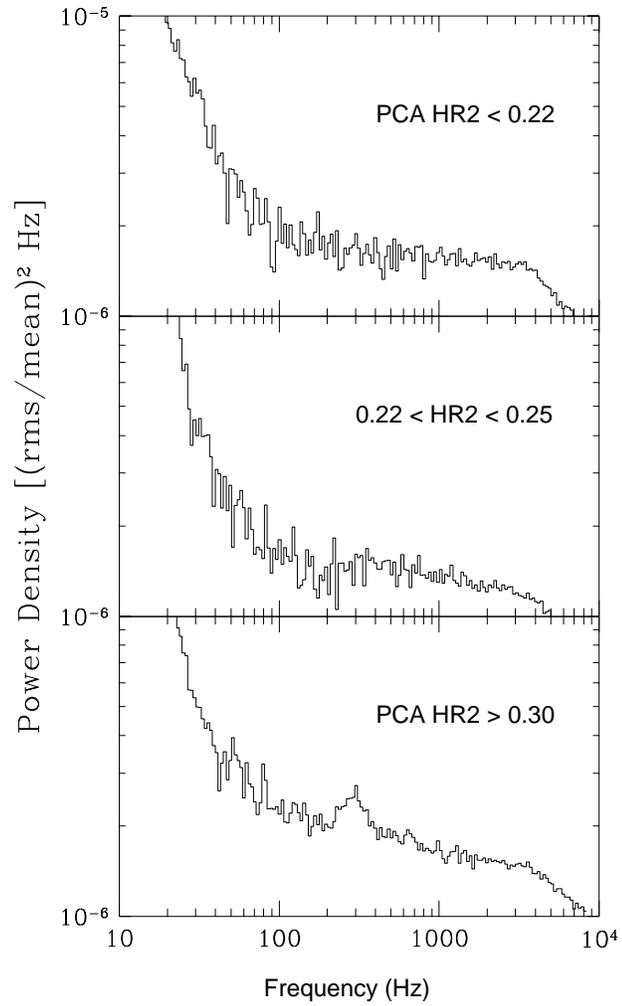,height=5.5in} 
\caption{PCA power spectra of GRO J1655-40 in which 20 observations are combined into 3 intervals of $PCA\_HR2$ (see text).  A QPO appears at 300 Hz during the 7 observations with the hardest X-ray spectra.
\label{fig:pds16b}}
\end{figure}

\end{document}